\documentclass[final,3p,times,twocolumn]{elsarticle}
\usepackage{preprintcover}  
\PreprintCoverPaperTitle{Evidence for the associated production of a $W$ boson and \\ a top quark in ATLAS at $\sqrt{s}=7$~\tev}  
\PreprintIdNumber{CERN-PH-EP-2012-117}
\PreprintCoverAbstract{
This Letter presents evidence for the associated production of a $W$~boson
and a top quark using \LUMI\ of $pp$ collision data at $\rm \sqrt{s}=7$~\tev\ accumulated
with the ATLAS detector at the LHC. The analysis is based on the selection of the dileptonic
final states with events featuring two isolated leptons, electron or muon, with
significant transverse missing momentum and at least one jet. An approach based
on boosted decision trees has been developed to improve the discrimination of
single top-quark \Wt\ events from background.
A template fit to the final classifier distributions is performed to determine
the cross-section. The result is incompatible with the background-only hypothesis
at the 3.3$\sigma$ level, the expected sensitivity assuming the Standard Model
production rate being 3.4$\sigma$.
The corresponding cross-section is determined and found to be
$\sigma_{\it{Wt}} = \rm 16.8\pm 2.9~(stat)\pm 4.9~(syst)~pb$,
 in good agreement with the Standard Model expectation. From
this result the CKM matrix element $|V_{\it{tb}}| = 1.03^{+0.16}_{-0.19}$ is derived
assuming that the $Wt$ production through $|V_{\it{ts}}|$ and $|V_{\it{td}}|$ is small.
}  
\PreprintJournalName{Physics Letters B}

\usepackage{multicol}
\usepackage{amssymb}
\usepackage{lineno}
\usepackage{atlasphysics}
\usepackage{subfigure}
\usepackage{mathrsfs}
\usepackage{extra_definitions}

\journal{Physics Letters B}

\begin{document}

\begin{frontmatter}

\title{Evidence for the associated production of a $W$ boson and a top quark \\ in ATLAS at $\sqrt{s}=7$~\tev}

\author{The ATLAS Collaboration}

\address{}

\begin{abstract}
This letter presents evidence for the associated production of a $W$~boson 
and a top quark using \LUMI\ of $pp$ collision data at $\rm \sqrt{s}=7$~\tev\ accumulated 
with the ATLAS detector at the LHC. The analysis is based on the selection of the dileptonic 
final states with events featuring two isolated leptons, electron or muon, with 
significant transverse missing momentum and at least one jet. An approach based 
on boosted decision trees has been developed to improve the discrimination of 
single top-quark \Wt\ events from background. 
A template fit to the final classifier distributions is performed to determine 
the cross-section. The result is incompatible with the background-only hypothesis 
at the 3.3$\sigma$ level, the expected sensitivity assuming the Standard Model 
production rate being 3.4$\sigma$.
The corresponding cross-section is determined and found to be
$\sigma_{\it{Wt}} = \rm 16.8\pm 2.9~(stat)\pm 4.9~(syst)~pb$, 
 in good agreement with the Standard Model expectation. From 
this result the CKM matrix element $|V_{\it{tb}}| = 1.03^{+0.16}_{-0.19}$ is derived 
assuming that the $Wt$ production through $|V_{\it{ts}}|$ and $|V_{\it{td}}|$ is small.

\end{abstract}

\begin{keyword}
ATLAS\sep Top quark\sep W+t\sep single top-quark 

\end{keyword}

\end{frontmatter}

\section{Introduction}

\par
The observation of single top-quark production was first reported by both
D0~\cite{SGTOP-OBS-D0} and CDF~\cite{SGTOP-OBS-CDF} experiments at the Tevatron. 
The observations by the two experiments are consistent with the Standard Model (SM) 
expectation for single top-quark production resulting from two mechanisms, the $t$-channel
and the $s$-channel, measured inclusively. The third SM single top-quark production 
mechanism, the associated production of a top quark and a $W$~boson, has not 
been observed at the Tevatron.

\begin{figure}[htbp]
\begin{center}
\includegraphics[width=6.5cm]{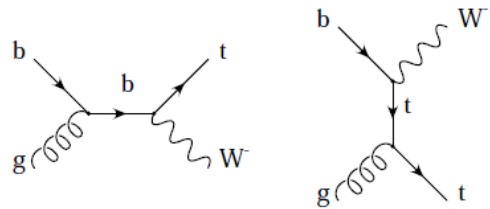}
\end{center}
\caption{{Leading-order Feynman diagrams for associated production of a single top-quark and a $W$ boson.}
\label{FIG-Feynman-Wt}}
\end{figure}

At the Large Hadron Collider (LHC), the electroweak production of single top-quarks
represents about half of the \TTB-pair production cross-section.  
First measurements of the single top-quark production~\cite{TCHAN-ATLAS,CMS-TCHAN}  
have been obtained in the \Tchan\ at a centre-of-mass energy of 7~\tev ,  
and show good agreement with the SM expectation. 
The associated production of a top quark and a $W$~boson involves the 
interaction of a gluon and a $b$-quark emitting an on-shell $W$~boson, as shown in 
the Feynman diagrams in Figure~\ref{FIG-Feynman-Wt}. The final state thus contains 
two $W$~bosons and an additional quark from the top quark decay, normally a $b$-quark.
Next-to-leading-order $Wt$ Feynman diagrams including a second $b$-quark may interfere with \TTB-pair production.
The interference should be small in the reconstructed exclusive final state with only one quark, 
where the largest fraction of $Wt$ signal is expected.
In this analysis, the $Wt$ leading-order approximation is used, and the difference
between leading-order and next-to-leading-order $Wt$ calculation is considered as modelling uncertainty.
Because of the massive particles in the final state, this production 
mechanism has an extremely low rate at the Tevatron compared to \Tchan , but is expected 
to have a much higher cross-section at the LHC, where the available 
partonic energy and the gluon flux are larger. For proton-proton collisions at 7~\tev, the 
single top-quark \Wtchan\ production cross-section is estimated to be 
$15.7\pm 1.1$~pb~\cite{XSECTION-WT} for a top quark mass of 172.5~\gev.

\par
Since the three modes of single top-quark production are sensitive to different 
manifestations of physics beyond the SM, measurements of the individual cross-sections 
are complementary to each other and allow some sources of new phenomena to be disentangled. 
The production mode with both a $W$ boson and a top quark in the final state has the special 
feature that both particles can be identified.  Thus, the measurement of the corresponding 
cross-section can be sensitive to new phenomena which modify the $W$-$t$-$b$ interaction, but 
insensitive to flavour-changing neutral currents (FCNCs) or new particles such 
as $W'$, $t'$ and techni-pions~\cite{MOTIVATION}.
 The measurement of the single top-quark \Wtchan\ production cross-sections 
therefore serves as a direct probe of the $W$-$t$-$b$ coupling and allows 
the direct determination of the quark-mixing matrix element $|V_{\it{tb}}|$~\cite{CKM1,CKM2}. 
This result can be compared to the results obtained from $t$- and $s$-channel 
production measurements. 

 In this Letter, an analysis is presented that establishes evidence 
for the associated production of a top quark and a $W$~boson in the dilepton 
channel, with $pp\to Wt \to \ell\nu b\ell\nu$, where $\ell=e,\mu$. 
Events featuring two leptons and neutrinos from $W$ boson decays and an additional  
jet originating from the top quark decay, are selected and analysed.
The corresponding cross-section is extracted and the magnitude of the 
CKM matrix element $|V_{\it{tb}}|$ is derived. Comparison is made  
with the Tevatron average and ATLAS measurements.

\section{Data and Monte Carlo simulation}

 The present analysis uses LHC proton-proton collision 
data at a centre-of-mass energy of 7~\tev\ collected between March and July 2011 with the ATLAS 
detector~\cite{ATL-2008-001}, which is composed of inner tracking detectors 
in a 2~tesla magnetic field surrounded by calorimeters and a muon spectrometer. The selected 
events were recorded based on single-electron or single-muon triggers. 
Detector and data-quality requirements are applied offline, 
resulting in a data set corresponding to an integrated luminosity of 
$\rm 2.05\pm 0.08~fb^{-1}$~\cite{ATL-LUMI,ATL-CONF-2011-116}.

In the following, all Monte Carlo (MC) simulations of top-quark related 
processes assume a top-quark mass of 172.5~\gev, and a width of 1.3~\gev, 
consistent with the world average value~\cite{Group:2011qkmass}. 
Samples of simulated events for single top-quark processes are produced 
with {\rm AcerMC} version~3.7~\cite{Kersevan:2004yg} coupled with 
the {\sc MRST2007}~\cite{Sherstnev:2007nd}   
parton distribution functions (PDFs). The \TTB-pair processes are generated 
using {\sc MC@NLO} version 3.41~\cite{SAMPLES-MCNLO}, 
interfaced with the {\sc CTEQ6.6} PDFs set~\cite{Nadolsky:2008zw}. 
All top quark samples are normalised using next-to-next-to-leading order (NNLO) 
cross-sections~\cite{XSECTION-WT,Moch:2008ai,langenfeld2009new,Beneke}.  
Gauge boson ($W/Z$) production in association with jets 
is simulated using the leading-order generator {\sc ALPGEN} version 2.13~\cite{SAMPLES-ALPGEN}, 
coupled with {\sc CTEQ6L1} PDFs~\cite{Lai:2010vv}. 
The diboson processes $WW$, $WZ$ and $ZZ$ are generated using {\sc ALPGEN} version 2.13 
with {\sc MRST2007} PDFs. 
In all cases, {\sc HERWIG}~\cite{SAMPLES-HERWIG} is used for the showering and is linked to the underlying event model in 
{\sc JIMMY} version 4.31~\cite{SAMPLES-JIMMY}. 
After the event generation, all samples are passed through the full simulation of the ATLAS 
detector~\cite{ATL-2010-005_sgtop} based on {\sc GEANT4}~\cite{SAMPLES-G4} and are 
reconstructed using the same procedure as collision data. The simulation 
includes the effect of a variable number of proton-proton collisions per bunch crossing and
is weighted to reproduce the same distribution of the number of collisions per bunch 
crossing as observed in data. The average number of interactions per bunch crossing is 6.2 in this 
data set.

\section{Event reconstruction and selection}

A set of general-purpose event-quality requirements~\cite{ATLAS-JETCLEAN} are applied to 
the data. Events are selected if they contain at least one primary vertex candidate 
with a minimum of five associated tracks, each reconstructed with transverse momentum  
(\pT) above 400~\mev. 
Events must not contain any jet, with \pT\ 
(calculated with the electromagnetic response for jets) 
greater than 20~\gev, arising from out-of-time energy depositions 
or from real energy depositions with a hardware or calibration problem.

\par
 Electron candidates are reconstructed using a cluster-based 
algorithm~\cite{ATL-ELEC-PERF} 
and are required to have transverse energy $\rm{\it E}_{\rm T}> 25~\gev$ and $|\eta|<2.47$, where $\eta$ denotes the
pseudorapidity\footnote{ATLAS uses a right-handed coordinate system with its origin at the nominal interaction point (IP) in 
the centre of the detector and the $z$-axis along the beam pipe. The $x$-axis points from the IP to the centre
of the LHC ring, and the $y$-axis points upwards. Cylindrical coordinates $(r, \phi)$ are 
used in the transverse plane, $\phi$ is the azimuthal angle around the beam pipe. 
The pseudorapidity $\eta$ is defined in terms of the polar angle $\theta$ as 
$\eta = -~\rm{ln}(\tan\theta/2)$.}. Events with electrons 
falling in the calorimeter barrel-endcap transition region, corresponding 
to $1.37<|\eta|<1.52$, are rejected. Candidates must satisfy a set of 
quality criteria, referred to as either ``loose'' or ``tight'' criteria~\cite{ATL-ELEC-PERF}, 
which for the latter, includes additional stringent requirements on the matching between the electron track candidate 
and the cluster. Isolation criteria require that the sum of the calorimeter transverse 
energy within a cone of radius $\Delta R=\sqrt{(\Delta\eta)^2+(\Delta\phi)^2}=0.3$ around the electron direction
(excluding the cells associated with the electron) must be less than 15\% of 
the electron transverse energy. In addition, the sum of the \pT\ of all tracks 
within the same cone radius around the electron direction, excluding the track 
belonging to the electron, must be less than 10\% of the electron \ET.

 Muon candidates are reconstructed by combining track segments found in the 
inner detector and in the muon spectrometer, and are required to have
$p_{\rm T} > 25~\gev$ and $|\eta|<2.5$. 
Selected muons must additionally satisfy a series of cuts on the number of hits on the track 
in the various tracking sub-detectors, referred to 
as ``tight'' quality criteria~\cite{ATL-2010-012-2}. The isolation requirements are the same 
as those for electrons. In order to reject events in which 
a muon emitting a hard photon is also reconstructed as an electron, events are vetoed when 
a selected electron-muon pair shares the same inner detector track. 

 Hadronic jets are reconstructed from calorimeter clusters~\cite{ATLAS-TOPO-CLUSTER} 
using the anti-$k_{t}$ algorithm~\cite{Salam:2008} with a radius parameter $R$ = 0.4.
To take into account the differences in calorimeter response to electrons 
and hadrons, a \pT- and $\eta$-dependent scale factor is 
applied to each jet in order to make an average energy scale correction~\cite{ATLAS-JES-2011}. 
Jets are required to have $E_{\rm T} > 30$~\gev\ and $|\eta| < 2.5$. 
Jets overlapping with selected electron candidates within $\Delta R<0.2$ are removed, 
keeping the electron candidate. 
The missing transverse momentum $\vec{E}_{\rm T}^{\rm miss}$ is calculated using the clusters 
identified in the calorimeter that are calibrated according to the associated reconstructed 
high-$p_{\rm T}$ objects. Taking also into account the energy clusters not associated to 
any high-$p_{\rm T}$ objects, projections of this vectorial sum in the transverse plane, 
correspond to the negative of the $\vec{E}_{\rm T}^{\rm miss}$ components. The missing 
transverse momentum is also corrected for the presence of electrons, muons, and 
jets~\cite{ATLAS-MET}.

\par
A dilepton event preselection classifies the events according to 
exclusive $ee$, $e\mu$ and $\mu\mu$ categories. 
The following event selections are common to all three $ee$, $\mu\mu$ and $e\mu$ 
channels. Candidate events 
must contain two ``tight'' opposite-sign leptons. Events having any additional isolated leptons
with \pt\ greater than 25~\gev\ are vetoed in order to ensure the orthogonality 
of the $ee$, $e\mu$ and $\mu\mu$ categories and suppress diboson backgrounds. 
Since the signal signature contains a single high-$p_{\rm T}$ quark from top quark decay, only events 
with at least one jet are selected. However, no $b$-tagging requirements are applied 
as they do not offer significant rejection over the primary background 
originating from \TTB-pair events. As signal events also 
feature neutrinos from the leptonic decays of $W$~bosons, the magnitude of the missing transverse 
momentum of the event is required to be greater than 50~\gev.

 In the $ee$ and $\mu\mu$ channels, the invariant mass of the lepton pair 
$m_{\ell\ell}$ is required to satisfy $m_{\ell\ell}<81$~\gev\ or $m_{\ell\ell}>101$~\gev\  
in order to reduce the contamination from $Z$~boson decays. In all three channels, the 
\Ztt\ background is reduced by applying a selection on the 
sum of the two angles in the transverse plane between each lepton and the missing transverse
momentum direction:

$$\Delta\phi(\ell_1,\vec{E}_{\rm T}^{\rm miss})+\Delta\phi(\ell_2,\vec{E}_{\rm T}^{\rm miss})>2.5~.$$
The application of this cut results in an expected rejection of 95\% of \Ztau\ 
events, 30\% of \Zee\ and \Zmm\ events and 21\% of \TTB-pair events, while 
keeping 87\% of the expected signal rate.
After the selection, signal is expected mainly in events with exactly 
one jet. Events with at least two jets are expected to be dominated 
by background events and are used as control regions.  

\section{Background estimation}

 The main background originates from \TTB-pair production in 
the dilepton channel $t\bar{t}\to\ell\nu b\ell\nu b$. 
The \TTB-pair background is estimated using MC simulation normalised 
to the NNLO cross-section~\cite{Moch:2008ai,langenfeld2009new,Beneke}, and the uncertainty is
further constrained by the fit of data in 2-jet and $\ge$3-jet bins.

Diboson events, where initial state radiation produces a jet that passes the jet selection requirements, 
represent about 15\% of the background in events selected with exactly one jet. 

Drell-Yan including $Z^{(*)}$ events can be selected if they contain an additional jet from gluon 
radiation. The contribution of the Drell-Yan process to the background in the 
$ee$ and $\mu\mu$ categories is determined via a data-driven 
procedure. In this method, orthogonal cuts on the reconstructed dilepton 
invariant mass $m_{\ell\ell}$ and the missing transverse momentum \MET\ variables are 
used to define a set of six regions, including two signal-enriched and four 
background-enriched regions for the $ee$ final state or the $\mu\mu$ final state. 
The contamination of the signal regions by Drell-Yan events is estimated from data 
which are scaled by the measured ratio of numbers of events selected in the 
corresponding control regions. 
This scale factor is corrected for the contamination by non-Drell-Yan backgrounds 
(top quark production, diboson, $W+\rm jets$) that are predicted by MC simulation 
and subtracted prior to its determination. Both the scale factor and non-Drell-Yan 
background-specific normalisation factors are determined using a likelihood fit 
of data in bins of $E_{\rm T}^{\rm miss}$. Variations by $\pm 1\sigma$ of these 
scale and normalisation factors are used to estimate the systematic uncertainty affecting the 
Drell-Yan event 
yield. The total uncertainty (statistical plus systematic) ranges between 10\% 
and 35\% depending upon the jet multiplicity. Drell-Yan events contribute about 
5\% of selected events. 

Contamination of selected events by ``fake dileptons'' may occur if a lepton from real W/Z decay and another lepton from
jet misidentification or heavy-flavour ($b$- and $c$-hadron) decays are selected, 
or both leptons from jet misidentification or heavy-flavour decays are selected,
such as \TTB-pair lepton+jets final state, $W$+jets or multijet events. 
These backgrounds are difficult to model accurately, so a data-driven approach 
based on the matrix method~\cite{Aad:2010ey} is followed. 
The method builds upon the use of ``tight'' and ``loose'' lepton selection 
criteria mentioned in Section~3. For these backgrounds, the efficiency for 
a ``loose'' lepton 
to be reconstructed as a ``tight'' lepton is determined 
using a data sample enriched in multijet events, where some of the 
lepton quality criteria have been reversed and the isolation requirement has been 
removed. The ``loose'' to ``tight'' efficiency for real leptons is measured from 
$Z\to \ell\ell$ events using a tag-and-probe analysis technique. 
The composition of the selected dilepton sample is extracted by inverting 
a $\rm 4\times 4$ matrix which relates the observed sample composition in 
terms of selected leptons of different quality to its true composition in 
terms of real and ``fake'' leptons. The background originating from these 
events represents less than 1\% of the selected sample. The corresponding 
systematic uncertainty is taken conservatively at 100\%. 

 A data-driven technique has been used to check the MC prediction of
the $Z\to\tau\tau$ contamination. The selected sample is split into
background- and signal-enriched regions, using the summed $\Delta\phi$ 
between the leptons and the $\vec{E}_{\rm T}^{\rm miss}$ direction requirement, as defined in Section~3.
The $Z\to\tau\tau$ background in the signal region is extracted using
the ratio of the corresponding MC estimates in both regions, scaled by the
number of selected data events from which non-Drell-Yan as well 
as Drell-Yan $ee$ and $\mu\mu$ backgrounds have been subtracted using MC. 
The difference between the purely MC-based expectations and this 
determination is included as a systematic error
and results in an uncertainty of 60\%. The $Z\to\tau\tau$ events constitute less 
than 1\% of the selected event sample. 

\begin{figure}[phtbp]
\centering
\subfigure[]{
\includegraphics[width=6.0cm]{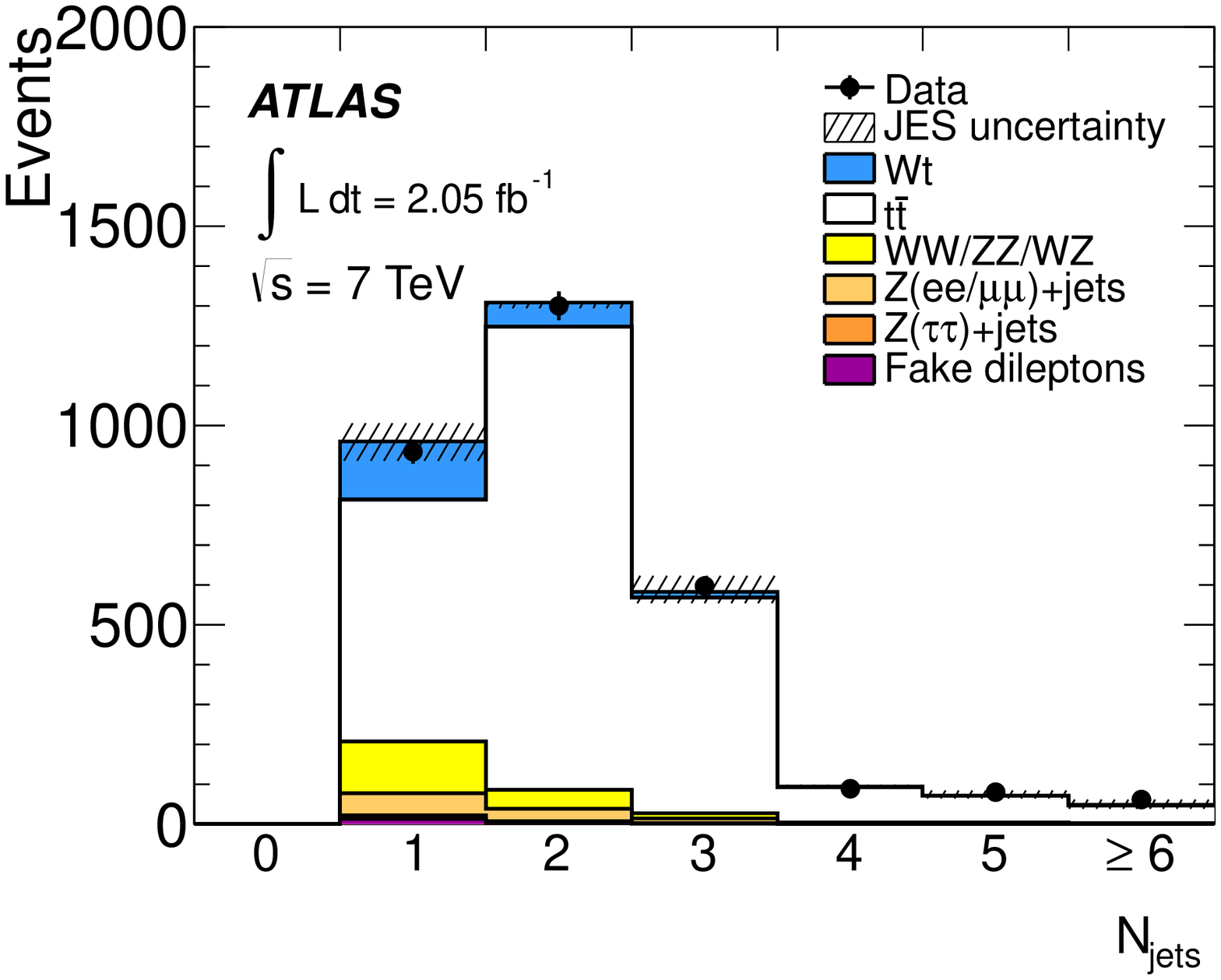}
\label{FIG-BDT-INPUT-a}
}
\subfigure[]{
\includegraphics[width=6.0cm]{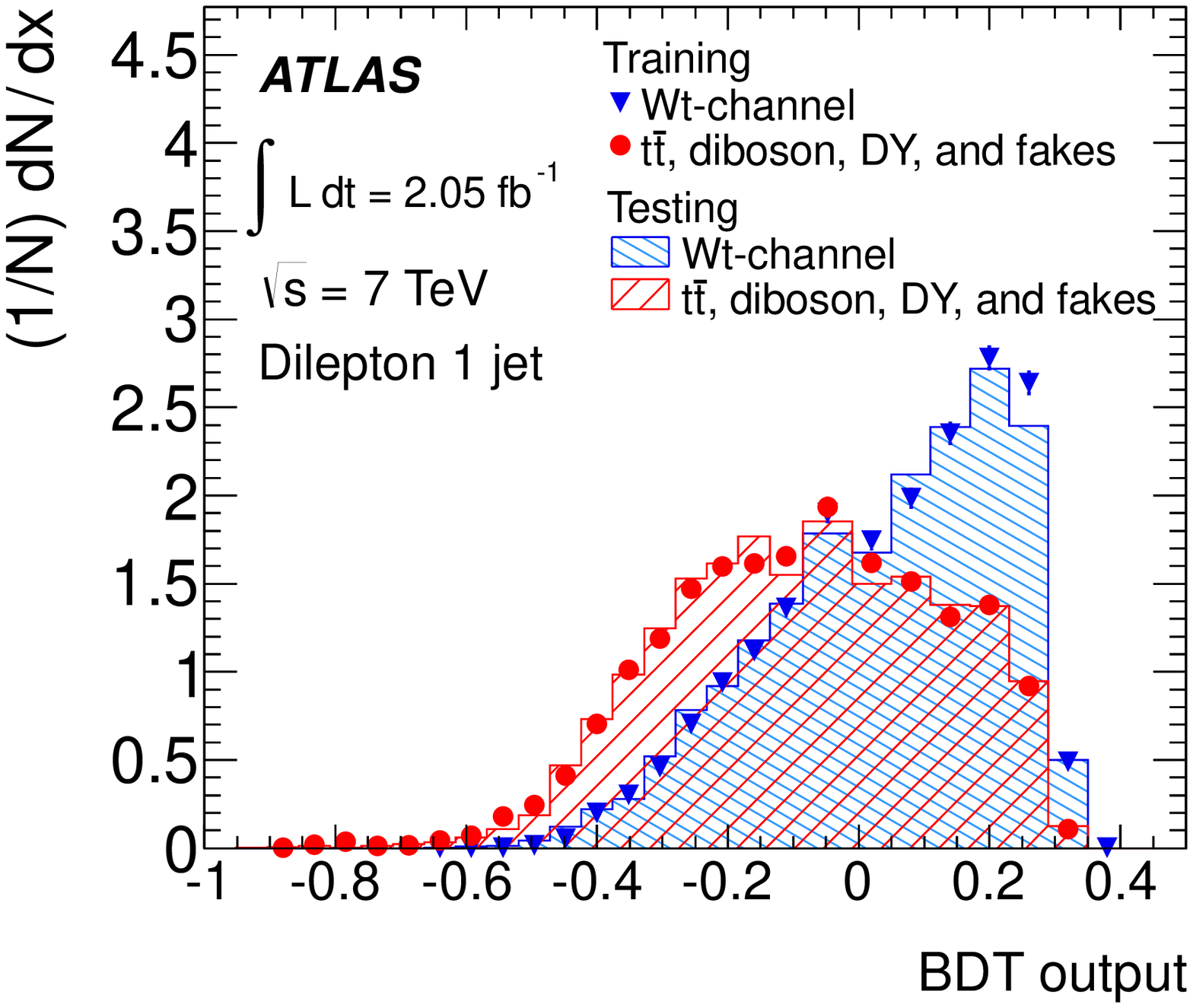}
\label{FIG-BDT-INPUT-b}
}
\caption{\subref{FIG-BDT-INPUT-a} Number of jets with $p_{\rm T}>30~$\gev\ and $|\eta|<2.5$ 
after the selection; hatched bands show the jet energy
scale (JES) uncertainty. The $Wt$ signal is normalised to the theory prediction. \subref{FIG-BDT-INPUT-b}
Distribution of BDT output for the signal (\Wtchan) and background (\TTB\, diboson, Drell-Yan and fake dileptons) in signal enriched 1-jet bin. 
The BDT method uses 2 statistically independent sets of MC-simulated events, indicated as training and testing samples,  
to check both signal and background BDT output stability. The BDT weight file is derived from a training sample and applied to a testing sample.} 
\end{figure}

The jet multiplicity distribution is shown in Figure~\ref{FIG-BDT-INPUT-a} 
after the selection described in Section 3. Table~\ref{TABLE-EVENT-YIELD-PRESEL} reports 
the expected signal, estimated backgrounds and total event yields in the 1-jet, 2-jet and $\ge$3-jet categories, with $ee$, $\mu\mu$
and $e\mu$ channels combined. 
No contamination from $t$-channel or $s$-channel single top-quark events is expected in the 
dilepton final state. A total of 224 signal events are expected over a 
background of 2840. The dominant \TTB-pair production accounts for 75\% of 
the background yield in 1-jet events.

\begin{table}[phtdp]
\begin{center}
   \begin{tabular}{l|rrr}
    \hline
              & 1-jet              & 2-jet             & $\rm \ge 3$-jet \\
    \hline \hline 
    $Wt$      &  $\rm 147\pm 13$   & $\rm 60\pm 9$    & $\rm 17\pm 5$ \\
    \hline
    \TTB\     &  $\rm 610\pm 110$  & $\rm 1160\pm140$  & $\rm 740\pm 130$ \\ 
    Diboson   &  $\rm 130\pm 17$   & $\rm 47\pm 5$    & $\rm   17\pm 4$ \\ 
    \Zee\     &  $\rm 20\pm  2$    & $\rm 11\pm 2$     & $\rm   5\pm 2$ \\ 
    \Zmm\     &  $\rm 29\pm  3$    & $\rm 28\pm 3$     & $\rm  12\pm 3$ \\ 
    \Ztt\     &  $\rm 9 \pm  6$    & $\rm  4\pm 3$     & $\rm   2\pm 1$ \\ 
    Fake dileptons & $\rm 11\pm 11$ & $\rm  5\pm 5$     & $\rm negl.$ \\ 
   \hline
 Total bkgd.  &   $\rm 810\pm 120$ & $\rm 1260\pm 140$ & $\rm 780\pm 130$ \\  
 Total expected   &   $\rm 960\pm 120$ & $\rm 1320\pm 140$ & $\rm 790\pm 130$ \\  
 \hline
 Data observed    &   $\rm 934$       & $\rm 1300$         & $\rm 825$ \\
    \hline\hline
   \end{tabular}
 \caption{Observed and expected event yield in the selected dilepton sample 
in the 1-jet, 2-jet and $\ge$3-jet bins for an integrated luminosity of 
\LUMI. The $Wt$, \TTB\ and diboson expectations are normalised to the theory predictions. 
Dilepton and lepton+jets channels are included in \TTB . Only
leptonic decays of diboson events are considered.
``Fake dileptons'' are events with at least one fake lepton, as described in the text. Uncertainties 
are the sum of statistical and systematic sources added in quadrature.}
\label{TABLE-EVENT-YIELD-PRESEL}
\end{center}
\end{table}
\noindent

\section{Discriminating variables for $Wt$ events}

 After the event selection, the signal-to-background ratio is 18\% 
in 1-jet events, where most of the signal is expected. As no individual variable 
is found to carry a large discriminating power, the analysis strategy uses a 
multivariate approach based on the ``boosted decision trees'' (BDT)~\cite{BDT-Breiman} 
technique in the framework of TMVA~\cite{Hocker:2007ht} to discriminate between the \Wtchan\ and 
\TTB-pair production. The BDT method benefits from the advantage of using the 
correlations between variables as part of the distinguishing power.
The goal is to exploit the differences between 
signal and background in many specific kinematic and topological distributions to 
form a classifier. This BDT classifier is trained using 1-jet events to maximise 
the expected significance without overtraining. BDT classifiers using the same input variables 
are also formed for 2-jet events and events with at least 3 jets: while no 
significant signal yield is expected in these events, the BDT output distribution serves to
constrain the background normalisation.    

Twenty-two variables with significant separation power 
are used as input to the BDT, all of which are well modelled by simulation.
The two most powerful variables are $p_{\rm T}^{\rm sys}$, defined as the 
magnitude of the vectorial sum of $p_{\rm T}$ of the leading jet, 
leptons and missing transverse momentum, and the ratio 
$p_{\rm T}^{\rm sys}/\sqrt{H_{\rm T}+\sum{E_{\rm T}}}$, where 
$H_{\rm T}$ is the scalar sum of the two leptons and the leading jet transverse momenta, and 
$\sum{E_{\rm T}}$ the scalar sum of the transverse energies of all energy deposits 
in the calorimeter.  
Other variables with lesser discriminating power are:
the event centrality, the thrust and its associated pseudorapidity,
the transverse momentum and pseudorapidity of the leading jet,
the pseudorapidity of each lepton,
the transverse momentum and pseudorapidity of the system formed by the dilepton 
and the leading jet, the invariant masses formed by each individual lepton 
with the leading jet, the missing transverse momentum, 
the azimuthal angle between the dilepton system and the leading jet 
directions, the pseudorapidity difference between the dilepton system and the leading 
jet, and the minimal azimuthal angle between the two leptons and the leading jet. 

\begin{figure}[htbp]
\centering
\subfigure[]{
\includegraphics[width=6.0cm]{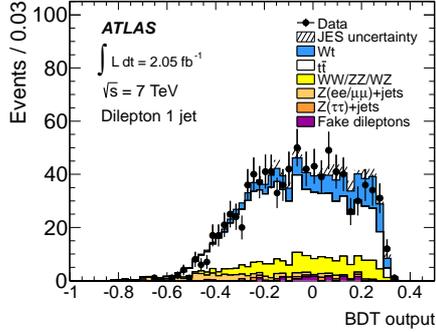}
\label{FIG-BDT-FINAL-a}
}
\subfigure[]{
\includegraphics[width=6.0cm]{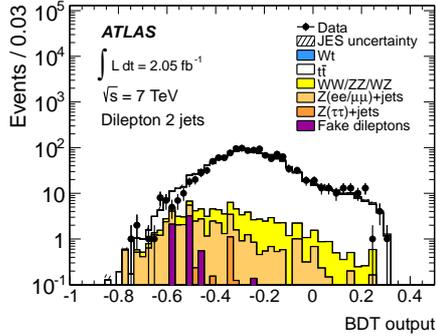}
\label{FIG-BDT-FINAL-b}
}
\subfigure[]{
\includegraphics[width=6.0cm]{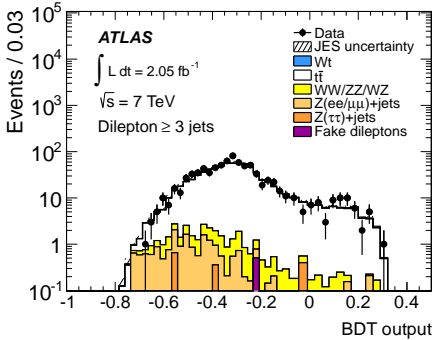}
\label{FIG-BDT-FINAL-c}
}
\label{FIG-BDT-FINAL}
\caption{BDT output for selected events in \subref{FIG-BDT-FINAL-a} 1-jet,   
\subref{FIG-BDT-FINAL-b} 2-jet and \subref{FIG-BDT-FINAL-c} $\ge$3-jet categories. The $Wt$ signal is normalised to the theory prediction in
all three categories.}
\end{figure}

Figure~\ref{FIG-BDT-INPUT-b} displays the BDT output probability density functions for signal 
and background in 1-jet events. Several checks are performed to ensure that the 
input variables are well modelled in a large phase space: both background-enriched 
regions, defined by events with exactly two jets and with at least three jets, 
and regions where most of the signal events are expected. Figures~\ref{FIG-BDT-FINAL-a}, 
\ref{FIG-BDT-FINAL-b} and ~\ref{FIG-BDT-FINAL-c} show the resulting good agreement 
of BDT outputs for data and MC simulation for 1-jet events, 2-jet events and events 
with at least 3 jets, respectively. 
\par

\section{Cross-section determination}

  In order to determine the cross-section, a template fit is performed 
to the three BDT output distributions for 1-jet, 2-jet and $\ge $3-jet 
events. 
 The determination of the \Wtchan\ single top-quark production 
yield is treated as a counting experiment in each bin 
and modelled using a likelihood function in terms of Poisson 
and Gaussian distributions:

$$
\mathcal{L}(\sigma_{Wt},\vec{\alpha}) = \prod_{i=1}^{3} \prod_{j=1}^{\rm{N_{bin}}}{\mathcal{P}}\left(N^{\rm{obs}}_{i,j} | N^{\rm{exp}}_{i,j}
(\vec{\alpha}) \right)\prod_{k=1}^{\rm{N_{syst}}} G(\alpha_{k}|0,1)
$$
where the index $i$ runs over the three jet multiplicity bins (1-jet, 2-jet and $\ge$3-jet), 
and $j$ runs over all bins of the corresponding BDT output distribution. The variables $N^{\rm{exp}}_{i,j}$ and 
$N^{\rm{obs}}_{i,j}$ are summed over the three dilepton flavour combinations. The index $k$ runs 
over the list of systematic uncertainty sources, which are presented below.

The likelihood function includes a Poisson term ${\mathcal{P}}\left(N^{\rm{obs}}_{i,j} | N^{\rm{exp}}_{i,j}(\vec{\alpha}) \right)$ in the observed 
number of events $N^{\rm obs}_{i,j}$ with the expectation value $N^{\rm exp}_{i,j}$ 
defined as the sum of the expected contributions from signal and all MC- or data-driven 
backgrounds in bin $j$ for the jet multiplicity bin $i$. 
Systematic uncertainties are grouped in uncorrelated sets ($k$) and their effect 
is parameterised for each $k$ using a nuisance 
parameter $\alpha_k$, where $\alpha_k=0$ maps to the nominal value and $\alpha_k = \pm 1$ 
map to $\pm 1\sigma$ shifts of the parameter. Piecewise-linear interpolation 
is used to propagate the effect of the $\alpha_k$ to the signal and background 
yields. A Gaussian shape $G(\alpha_k|0,1)$ centred at zero with unit
width is used for the $\alpha_k$ constraint terms in the likelihood. 

The contributions to the uncertainty on the fitted \Wtchan\ cross-section are 
shown in Table~\ref{TABLE-XS-UNCERTAINTIES} and further described below. 
The main experimental 
source of systematic uncertainties comes from the knowledge of the 
jet energy scale (JES), which carries an uncertainty of 2\% to 7\% parameterised 
as a function of jet $p_{\rm T}$ and $\eta$~\cite{ATLAS-JES-2011}. The presence of a $b$-jet in 
the event is also taken into account and an extra uncertainty of 2\% to 5\% depending on jet 
\pt\ is added in quadrature to the non-$b$-jet uncertainty. Other experimental 
uncertainty sources which have been considered are the jet energy resolution, the jet 
reconstruction efficiency, the lepton identification efficiency, the lepton 
energy scale determination and resolution as well as the multiple proton-proton 
collision and underlying event modelling. The uncertainty in the 
luminosity determination is 3.7\%~\cite{ATL-LUMI,ATL-CONF-2011-116}.

 Uncertainties in the simulation include the effects of the MC 
generator choice, the scheme used in the hadronisation and showering and   
models of the initial and final state radiation (ISR/FSR). 
Generator choice uncertainty is estimated by comparing 
{\rm AcerMC} with {\sc MC@NLO} generators for single top-quark $Wt$ events,
and comparing {\sc POWHEG} with {\sc MC@NLO} generators for top quark pair events.
Hadronisation and showering effects are estimated using the differences seen 
in generated events interfaced with either {\sc PYTHIA}~\cite{Sjostrand:2006zamy} or {\sc HERWIG}. 
Finally, ISR/FSR modelling effects are assessed on MC signal and background 
samples interfaced with {\sc PYTHIA}. Specific tunes are used to 
separately vary ISR and FSR modelling via 
changes to $1/\Lambda^{\rm ISR}_{\rm QCD}$, the maximum parton virtuality 
in a space-like parton shower, the $\Lambda^{\rm FSR}_{\rm QCD}$ scale and the 
FSR infrared cut-off~\cite{ATL-ISR-FSR}. 

  The impacts on both acceptance and kinematic distributions
shapes are considered for the experimental and simulation uncertainties.

 Remaining theoretical uncertainty sources include the cross-section normalisation for 
the \TTB-pair background ($^{+7\%}_{-10\%}$)~\cite{Moch:2008ai,langenfeld2009new,Beneke} 
and diboson production ($\pm 5\%$)~\cite{Aad:2010ey}, as well as the choice of the 
parton distribution functions. For the latter, acceptance variations have been assessed using the
CTEQ~\cite{Lai:2010vv}, 
MRST~\cite{springerlink:10} and NNPDF~\cite{Ball:2010de} sets. 

 The cross-section is obtained by maximising the likelihood function 
using {\sc RooFit}~\cite{ROOFIT}. The total uncertainty is inferred from the shape of 
the profile likelihood ratio~\cite{ATLAS-TEST-STATISTIC}:

$$\rm -2 ln\frac{\mathcal{L}(data|\sigma_{\it{Wt}},\hat{\vec{\alpha}}_{\sigma_{\it{Wt}}})}{\mathcal{L}(data|\hat{\sigma}_{\it{Wt}},\hat{\vec{\alpha}})}~,$$
where $\hat{\vec{\alpha}}$ and $\hat{\sigma}_{\it{Wt}}$ are the parameters that maximise the likelihood with the constraint of
$\hat{\sigma}_{\it{Wt}}>0$, and $\hat{\vec{\alpha}}_{\sigma_{\it{Wt}}}$ are the nuisance parameter values that maximise the likelihood for a given
$\sigma_{\it{Wt}}$. The maximisation is performed by varying
all the nuisance parameters, except the systematic uncertainties due to the generator 
and the parton shower whose effects are estimated separately using pseudo-experiments.

 The inclusion of 2-jet and $\ge$3-jet events in the fit brings 
additional constraints on the effect of systematic uncertainties, 
as jet energy scale and resolution effects as well as ISR/FSR modelling directly
affect the jet multiplicity distributions and the BDT outputs. 
These effects have been evaluated by varying the corresponding nuisance parameter 
central values in the fit to the data. The studies show
that the fitted result for the cross-section is not biased by the models used to 
describe the JES and ISR/FSR uncertainties. 

\begin{table}[htdp]
\begin{center}
   \begin{tabular}{l|rr}
    \hline
    Source & \multicolumn{2}{c}{$\Delta\sigma_{\it{Wt}}/\sigma_{\it{Wt}}$ [\%]}\\
           & observed & expected \\
    \hline \hline

   Data statistics                   & $17$   & $17$  \\
   MC statistics                     & $<5$      & $<5$     \\
   \hline
   Lepton energy scale/res.          & $<5$      & $<5$     \\
   Lepton efficiencies               & $7$       & $6$    \\
   Jet energy scale                  & $16$      & $14$ \\
   Jet energy resolution             & $<5$      & $<5$     \\
   Jet reconstruction eff.           & $<5$      & $<5$     \\
   Generator                         & $10$      & $12$    \\
   Parton shower                     & $15$      & $14$  \\
   ISR/FSR                           & $5$       & $6$     \\
   PDF                               & $<5$      & $6$     \\
   Pile-up                           & $10$      & $7$   \\
   $t\bar{t}$ cross-section          & $6$       & $6$    \\
   Diboson cross-section             & $6$       & $5$     \\
   Drell-Yan estimate                & $<5$      & $<5$     \\
   Fake dileptons estimate            & $<5$      & $<5$     \\
   $Z\to\tau\tau$ estimate           & $<5$      & $<5$     \\
   Luminosity                        & $7$     & $7$    \\
   \hline
   All systematics                   & $29$   & $29$  \\
   \hline\hline
   Total                             & $34$   & $33$  \\

    \hline\hline
   \end{tabular}
 \caption{Contributions to the uncertainty on the \Wtchan\ cross-section. The expected 
results assume the SM cross-section for the signal.}
\label{TABLE-XS-UNCERTAINTIES}
\end{center}
\end{table}

 The fitted result for the \Wt\ cross-section at 7~\tev\ is:

$$\sigma_{\it{Wt}} = \rm 16.8\pm 2.9~(stat)\pm 4.9~(syst)~pb.$$

In order to determine the sensitivity of the analysis, an ensemble test is
performed on pseudo-experiments. Systematic uncertainties are
treated as nuisance parameters which are constrained using Gaussian functions.
Both ``background-only'' and ``signal+background'' (where the signal rate is 
predicted by the SM) hypotheses are tested via the generation of dedicated sets 
of pseudo-experiments. The likelihood ratio defined as

$$\rm LLR = -2ln\frac{\mathcal{L}(data~|\sigma_{\it{Wt}}^{\rm SM},~\hat{\vec{\alpha}}_{\sigma_{\it{Wt}}^{\rm SM}})}{\mathcal{L}(data
|0,~\hat{\vec{\alpha}}_0)}$$
is computed for each pseudo-experiment. 
It is used to derive the $p$-value, which measures the probability for the 
background to fluctuate above the observed or expected number of events. 
This $p$-value is in turn 
interpreted in terms of significance and corresponds to 
a $3.3\sigma$ effect for the data. The corresponding significance for the
expected value assuming the SM cross-section corresponds to 
a $3.4\sigma$ effect.

\section{Determination  of $|V_{\it{tb}}|$}

 A direct determination of $|V_{\it{tb}}|$ can be extracted from the 
cross-section, assuming that the $Wt$ production through $|V_{\it{ts}}|$ and $|V_{\it{td}}|$ is small.
The \TTB\ background, which is the only background in the analysis that involves $|V_{\it{tb}}|^{2}$, 
does not affect this determination since top quark decays to a fourth generation heavier quark is disfavoured
by kinematics. 
The observed $|V_{\it{tb}}|^{2}$ is obtained by dividing the measured 
cross-section by the theoretical single top-quark cross-section calculated 
with a top quark mass of 172.5~\gev. 
Using  $\rm \sigma_{Wt}^{theory} = 15.7(\pm 1.1)\times |V_{\it{tb}}|^{2}~pb$~\cite{XSECTION-WT}, 
the following value is obtained for $|V_{\it{tb}}|$: 

$$|V_{\it{tb}}| = 1.03^{+0.16}_{-0.19}~,$$
 where the uncertainties in the cross-section measurement and in the theoretical 
predictions have been added in quadrature. This result is compatible with the 
combination of direct measurements at the Tevatron~\cite{Group:2009qk}:
$|V_{\it{tb}}| = 0.88^{+0.07}_{-0.07},$ and the measurement by ATLAS~\cite{TCHAN-ATLAS}: $|V_{\it{tb}}| = 1.13^{+0.14}_{-0.13}.$

\section{Conclusion}

 Evidence for the production of single top-quark events in 
the \Wtchan\ is reported with \LUMI\ of data collected at 7~\tev\ with ATLAS during 2011. 
The strategy followed consists of selecting dilepton events with at least one 
central jet. Drell-Yan and fake-dilepton backgrounds are estimated in data, while a 
classifier is used to optimise the discrimination of signal and \TTB-pair 
events. A fit of the classifier distributions is performed to extract 
the \Wtchan\ cross-section. The observed significance is 3.3 standard 
deviations for an expected sensitivity of 3.4. The corresponding fitted 
cross-section is 
$\sigma(pp\rightarrow Wt+X) = \rm 16.8\pm 2.9~(stat)\pm 4.9~(syst)$~pb.
A direct determination of $|V_{\it{tb}}|=1.03^{+0.16}_{-0.19}$ is extracted 
assuming that the $Wt$ production through $|V_{\it{ts}}|$ and $|V_{\it{td}}|$ is small.

\section{Acknowledgements}

We thank CERN for the very successful operation of the LHC, as well as the
support staff from our institutions without whom ATLAS could not be
operated efficiently.

We acknowledge the support of ANPCyT, Argentina; YerPhI, Armenia; ARC,
Australia; BMWF, Austria; ANAS, Azerbaijan; SSTC, Belarus; CNPq and FAPESP,
Brazil; NSERC, NRC and CFI, Canada; CERN; CONICYT, Chile; CAS, MOST and NSFC,
China; COLCIENCIAS, Colombia; MSMT CR, MPO CR and VSC CR, Czech Republic;
DNRF, DNSRC and Lundbeck Foundation, Denmark; EPLANET and ERC, European Union;
IN2P3-CNRS, CEA-DSM/IRFU, France; GNAS, Georgia; BMBF, DFG, HGF, MPG and AvH
Foundation, Germany; GSRT, Greece; ISF, MINERVA, GIF, DIP and Benoziyo Center,
Israel; INFN, Italy; MEXT and JSPS, Japan; CNRST, Morocco; FOM and NWO,
Netherlands; RCN, Norway; MNiSW, Poland; GRICES and FCT, Portugal; MERYS
(MECTS), Romania; MES of Russia and ROSATOM, Russian Federation; JINR; MSTD,
Serbia; MSSR, Slovakia; ARRS and MVZT, Slovenia; DST/NRF, South Africa;
MICINN, Spain; SRC and Wallenberg Foundation, Sweden; SER, SNSF and Cantons of
Bern and Geneva, Switzerland; NSC, Taiwan; TAEK, Turkey; STFC, the Royal
Society and Leverhulme Trust, United Kingdom; DOE and NSF, United States of
America.

The crucial computing support from all WLCG partners is acknowledged
gratefully, in particular from CERN and the ATLAS Tier-1 facilities at
TRIUMF (Canada), NDGF (Denmark, Norway, Sweden), CC-IN2P3 (France),
KIT/GridKA (Germany), INFN-CNAF (Italy), NL-T1 (Netherlands), PIC (Spain),
ASGC (Taiwan), RAL (UK) and BNL (USA) and in the Tier-2 facilities
worldwide.
\noindent

\bibliographystyle{atlasnote}
\bibliography{dilepton}
\onecolumn
\clearpage 

\begin{flushleft}
{\Large The ATLAS Collaboration}

\bigskip

G.~Aad$^{\rm 48}$,
B.~Abbott$^{\rm 111}$,
J.~Abdallah$^{\rm 11}$,
S.~Abdel~Khalek$^{\rm 115}$,
A.A.~Abdelalim$^{\rm 49}$,
O.~Abdinov$^{\rm 10}$,
B.~Abi$^{\rm 112}$,
M.~Abolins$^{\rm 88}$,
O.S.~AbouZeid$^{\rm 158}$,
H.~Abramowicz$^{\rm 153}$
% ...
\bigskip

% ...
$^{10}$ Institute of Physics, Azerbaijan Academy of Sciences, Baku, Azerbaijan\\
$^{11}$ Institut de F\'isica d'Altes Energies and Departament de F\'isica de la Universitat Aut\`onoma  de Barcelona and ICREA, Barcelona, Spain\\
% ...
$^{48}$ Fakult\"{a}t f\"{u}r Mathematik und Physik, Albert-Ludwigs-Universit\"{a}t, Freiburg i.Br., Germany\\
$^{49}$ Section de Physique, Universit\'e de Gen\`eve, Geneva, Switzerland\\
% ...
$^{88}$ Department of Physics and Astronomy, Michigan State University, East Lansing MI, United States of America\\
% ...
$^{111}$ Homer L. Dodge Department of Physics and Astronomy, University of Oklahoma, Norman OK, United States of America\\
$^{112}$ Department of Physics, Oklahoma State University, Stillwater OK, United States of America\\
% ...
$^{115}$ LAL, Universit\'e Paris-Sud and CNRS/IN2P3, Orsay, France\\
% ...
$^{153}$ Raymond and Beverly Sackler School of Physics and Astronomy, Tel Aviv University, Tel Aviv, Israel\\
% ...
$^{158}$ Department of Physics, University of Toronto, Toronto ON, Canada\\
% ...
\end{flushleft}


\end{document}